\documentclass[12pt]{article}
\usepackage{amssymb}
\usepackage{graphicx}
\usepackage{axodraw}
\def\be{\begin{equation}}
\def\ee{\end{equation}}

\begin{document}
\begin{center}
\hfill \vbox{
\hbox{December 2006}}
\vskip 0.8cm
{\Large\bf The spectrum of massive excitations of 3$d$ 3-state Potts model
and universality}\\ 
\end{center}
\vskip 0.3cm
\centerline{R. Falcone$^*$, R. Fiore$^*$, M. Gravina$^*$ and
A. Papa$^*$}
\vskip 0.3cm
\centerline{\sl Dipartimento di Fisica, Universit\`a della Calabria,}
\centerline{\sl and Istituto Nazionale di Fisica Nucleare, Gruppo collegato
di Cosenza}
\centerline{\sl I--87036 Arcavacata di Rende, Cosenza, Italy}
\vskip 0.6cm
\begin{abstract}
We consider the mass spectrum of the 3$d$ 3-state Potts model in the broken
phase (a) near the second order Ising critical point in the temperature - magnetic 
field plane and (b) near the weakly first order transition point at zero magnetic 
field. In the case (a), we compare the mass spectrum with the prediction from
universality of mass ratios in the 3$d$ Ising class; in the case (b),
we determine a mass ratio to be compared with the corresponding one in the 
spectrum of screening masses of the (3+1)$d$ SU(3) pure gauge theory at finite 
temperature in the deconfined phase near the transition. The agreement in
the comparison in the case (a) would represent a non-trivial test of
validity of the conjecture of spectrum universality. A positive answer
to the comparison in the case (b) would suggest the possibility to extend this
conjecture to weakly first order phase transitions.
\end{abstract}

\vfill
\hrule
\vspace{0.3cm}
\noindent$^*$ {\it e-mail addresses}: 
rfalcone,fiore,gravina,papa@cs.infn.it 

\newpage

\section{Introduction}

Universality is a powerful concept since it establishes the common 
long-distance behavior of theories characterized by different microscopic 
interactions, but possessing the same underlying global symmetry. 
A remarkable example is provided
by $(d+1)$-dimensional SU(N) pure gauge theories at finite temperature 
which undergo a confinement/deconfinement phase transition associated with the 
breaking of the center of the gauge group, Z(N), the order parameter being the
Polyakov loop~\cite{Polyakov:vu,Susskind:up}. When the transition is
second order, the long-range critical behavior of $(d+1)$-dimensional SU(N) pure 
gauge theories at finite temperature is conjectured to be the 
same as the $d$-dimensional Z(N) spin model in the critical region 
near the order/disorder phase transition~\cite{Svetitsky:1982gs}. 
As a consequence, the gauge theory and the spin model are predicted to have 
the same critical indices, amplitude ratios and correlation 
functions at criticality. This prediction has been accurately verified 
in several cases - see, for instance, 
Refs.~\cite{Gliozzi:1997yc,Fiore:1998uk,Engels:1998nv,Fortunato:2000hg,
Fiore:2001ci,Fiore:2001pf,Papa:2002gt} and, for a review, 
Ref.~\cite{Pelissetto-Vicari:2002}. 

A few years ago a study of the broken symmetry phase of the
3$d$ Z(2) (Ising) class has brought compelling evidence that universality
has a much wider reach than usually expected. In particular, it has been
shown that the Ising model and the lattice regularized $\phi^4$ 
theory both exhibit a rich spectrum of massive excitations and that
mass ratios coincide in the scaling region~\cite{Caselle:1999tm,Caselle:2001im}.
This result is quite far from obvious: since only the lowest mass 
contributes to the free energy, there is no simple reason why higher 
masses in the spectrum should be universal. Later on, numerical evidence has 
been given that the same spectrum characterizes (3+1)$d$ SU(2) gauge theory 
at finite temperature in the scaling region above the deconfinement 
temperature~\cite{Fiore:2002fj,Fiore-Papa-Provero-2003}. In the deconfined 
phase of the SU(2) gauge theory the counterpart of the massive excitations in 
the broken phase of the 3$d$ Ising model are the so-called {\em screening 
masses}, i.e. the inverse exponential decay lengths of correlation functions 
between suitably defined operators, built from the Polyakov 
loops~\cite{Fiore:2002fj,Fiore-Papa-Provero-2003}.

It would be quite interesting to extend the test of the universality of the
spectrum in two different directions: (1) by considering a theory with
a global symmetry different from Z(2), which, however, presents in its phase
diagram a second order critical point in the 3$d$ Ising class; (2) by verifying
if and to what extent the universality of the spectrum holds also in the
broken symmetry phase near a {\em weak} first order phase transition.

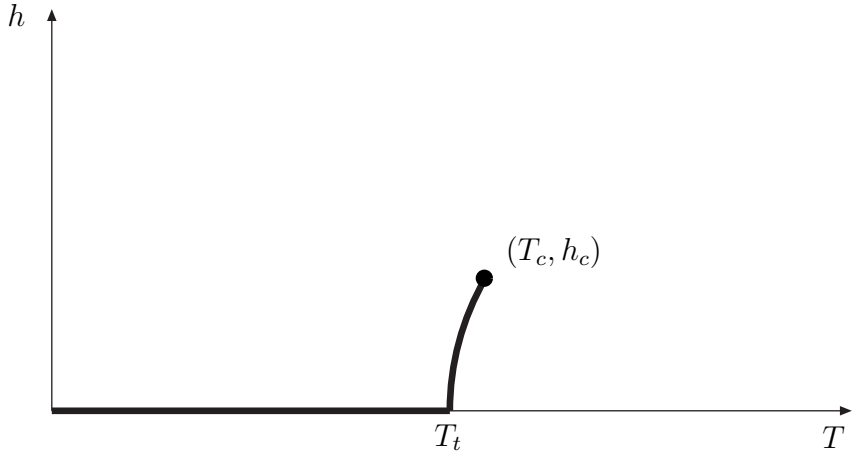
\begin{figure}[tb]
\centering
\label{phase_diag}
\begin{picture}(300,160)(0,0)

\LongArrow(0,10)(300,10)
\LongArrow(0,10)(0,160)
\SetWidth{2.5}
\Line(0,10)(150,10)
\CArc(250,10)(100,150,180)
\CCirc(163,60){2}{1}{1}
\SetWidth{0.5}

\Text(300,0)[r]{$T$}
\Text(-10,160)[r]{$h$}
\Text(150,0)[c]{$T_t$}
\Text(190,70)[c]{$(T_c,h_c)$}

\end{picture}
\caption[]{Qualitative phase diagram of the 3$d$ 3-state Potts model:
the solid line in bold is the line of first order phase transitions.
$T_t$ is the order/disorder transition temperature at zero magnetic field, 
$(T_c,h_c)$ is the endpoint in the 3$d$ Ising class.}
\end{figure}

The 3$d$ 3-state Potts model with external magnetic field model provides 
a good test-field for both these investigations. Indeed, its phase diagram 
in the temperature - magnetic field $(T,h)$ plane (see Fig.~\ref{phase_diag}) 
exhibits a line of first order phase transitions which starts at 
$(0,0)$, moves along the $T$-axis up to a transition temperature $T_t$ and then
bends in the positive $h$-plane up to reaching a second order endpoint, which
belongs to the Ising class~\cite{Karsch-Stickan}. The 3$d$ 3-state Potts model
near this critical endpoint should exhibit a mass spectrum in the Ising 
universality class. Instead, the mass spectrum of the 3$d$ 3-state Potts model
at zero magnetic field, in the broken phase near $T_t$, i.e. for $T\lesssim T_t$
should reproduce the spectrum of screening masses of the finite temperature
$(3+1)d$ SU(3) pure gauge theory in the deconfined phase near criticality, {\em if} 
universality holds also in the case of weak first order phase transitions.

In this paper we determine the low-lying masses of the spectrum of the 
3$d$ 3-state Potts model in two different sectors of parity and orbital 
angular momentum, 0$^+$ and 2$^+$, and in two different regions of the 
$(T,h)$ phase diagram, (a) near the critical Ising endpoint, 
(b) near the transition temperature $T_t$ at zero magnetic field. 
For the determinations in the region (a), the comparison between the 
resulting mass ratios and the corresponding ones in the 3$d$ Ising 
model~\cite{Caselle:1999tm} will provide a check of validity of the 
Ansatz of universality of the spectrum.
The mass ratios determined in the region (b) are instead the quantities to be
compared with the corresponding screening masses in the finite
temperature SU(3) gauge theory in order to verify the validity of the Ansatz 
near a weakly first order transition point.

This paper is organized as follows: in Section~\ref{Potts} we briefly review
the 3$d$ 3-state Potts model and describe its phase diagram on the temperature - 
magnetic field plane; in Section~\ref{sec:masses} we describe the method used to 
extract the massive excitations in the two regions of the phase diagram 
in which we are interested; in Section~\ref{results} we present our numerical 
results and in the last Section we draw our conclusions.

\section{The 3$d$ 3-state Potts model}
\label{Potts}

The 3-$d$ 3-state Potts model~\cite{Blote-Swendsen,Janke-Villanova} is a spin 
theory in which the fundamental degree of freedom, $s_i$, defined in the site $i$ of a 
3-dimensional lattice, is an element of the Z(3) group, {\it i.e.}
\be
s_i=e^{i\frac{2}{3}\pi\sigma_i}, \quad  \sigma_i=\{0,1,2\}\quad .
\ee
The Hamiltonian of the model is
\be
H=-\frac{2}{3}\beta \sum_{\langle i j \rangle} \big( s_i^\dagger s_j 
+ s_j^\dagger s_i \big) =-\beta \sum_{\langle i j \rangle} \delta_{\sigma_i,\sigma_j}\;,
\label{ham}
\ee
up to an irrelevant constant. Here, $\beta$ is the coupling in units of the temperature and the sum is done
over all the nearest-neighbor pairs of a cubic lattice with linear size $L$.

The Hamiltonian~(\ref{ham}) is invariant under the Z(3) transformation
\be
s_i \longrightarrow s_i^\prime=e^{i\frac{2\pi}{3}\sigma} s_i\;,
\ee
where $\sigma$ is fixed at any of the values $\{0,1,2\}$.
It is well known that this system undergoes a {\it weakly} first order
phase transition~\cite{Gavai-Karsch-Petersson}, associated with the 
spontaneous breaking of the Z(3) symmetry. The order parameter of this
transition is the magnetization, 
\be
\langle S \rangle=\langle \frac{1}{L^3}\sum_i s_i \rangle \;,
\ee
which, in the thermodynamic limit, is zero above the transition temperature 
$T_t$ (or below the transition coupling $\beta_t$) and takes a non-zero value 
below $T_t$ (or above $\beta_t$).

\begin{figure}[tb]
\centering
\label{phase_diag_xitau}
\begin{picture}(250,160)(0,0)

\LongArrow(0,10)(250,10)
\LongArrow(0,10)(0,160)
\SetWidth{2.5}
\Line(100,10)(250,10)
\CArc(300,100)(219.3,190,204)
\CCirc(84,62){2}{1}{1}
\SetWidth{0.5}
\LongArrow(70,140)(95,0)
\LongArrow(150,75)(10,49)

\Text(250,0)[r]{$\beta$}
\Text(-10,160)[r]{$h$}
\Text(110,20)[c]{$\beta_t$}
\Text(60,75)[c]{$(\tau_c,\xi_c)$}
\Text(10,60)[c]{$\xi$}
\Text(85,0)[c]{$\tau$}

\end{picture}
\caption[]{Qualitative phase diagram of the 3$d$ 3-state Potts model
in the $(\beta,h)$-plane: $\xi$ and $\tau$ are the symmetry-breaking and the 
temperature parameters of the Ising theory~(\ref{ising}).}
\end{figure}
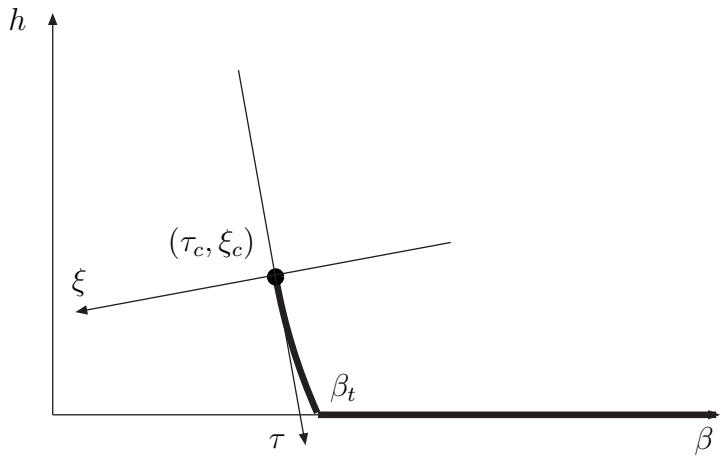

In presence of an external magnetic field it is convenient to work with an
Hamiltonian written in terms of the $\sigma_i$ degrees of freedom, instead of 
the $s_i$ ones. For a uniform magnetic field along the direction $\sigma_h$
with strength $h$ in units of the temperature, the Hamiltonian is
\be
H = -\beta \sum_{\langle i j \rangle} \delta_{\sigma_i,\sigma_j} 
-h \sum_i \delta_{\sigma_i,\sigma_h} \equiv -\beta E -h M \;,
\label{ham_magn}
\ee
where $E$ is the internal energy and $M$ is the magnetization.
The magnetic field breaks explicitly the Z(3) symmetry. However, first order 
transitions still occur for values of the magnetic field strength $h$ below a 
critical value $h_c$, the transition temperature increasing with increasing $h$. 
The line of first order phase transitions ends in a second order critical point 
$P_c=(T_c,h_c)$ (see Fig.~\ref{phase_diag}), belonging to the 3$d$ Ising 
class~\cite{Karsch-Stickan}. The Hamiltonian in the scaling region 
of the critical point $P_c$ can be written as 
\be
H=-\tau \tilde{E} -\xi \tilde{M} \;,
\label{ising}
\ee
where $\tilde E$ and $\tilde M$ are the Ising-like energy and magnetization operators
and $\tau$ and $\xi$ the corresponding temperature-like and symmetry-breaking-like 
parameters. This means that $\langle \tilde M \rangle$ is the new order parameter.
Close enough to $P_c$, the following relations hold,
\be
\tilde{M}=M+sE \;, \;\;\;\;\; \tilde{E}=E+rM \;,
\label{mixing}
\ee
where the mixing parameters $(r,s)$ have been determined numerically for several 
lattice sizes $L$ in Ref.~\cite{Karsch-Stickan}. The $\tau$-direction identifies 
the first order line (see Fig.~\ref{phase_diag_xitau}). In 
Ref.~\cite{Karsch-Stickan} the location of the critical point $P_c$ has been 
accurately determined: 
$P_c=(\beta_c,h_c)$=(0.54938(2),0.000775(10)) or, equivalently, 
$P_c=(\tau_c,\xi_c)$=(0.37182(2),0.25733(2)).

\section{Massive excitations and Universality}
\label{sec:masses}

Among the quantities relevant in the description of a phase transition there is
the correlation function of the local order parameter: in the case of the 3$d$ 
3-state Potts model this is just the local spin $s_i$. The point-point correlation
function is defined as
\be
\Gamma_{i_0}(r)=\langle s_i^\dagger s_{i_0} \rangle -\langle s_i^\dagger \rangle 
\langle s_{i_0} \rangle \;,
\ee
where $i$ and $i_0$ are the indices of the sites and $r$ is the distance between 
them. 
The large-$r$ behavior of the point-point correlation function is determined by
the correlation length of the theory, $\xi_0$, or, equivalently, by
its inverse, the fundamental mass. In order to extract the fundamental 
mass it is convenient to define the wall-wall correlation function, since 
numerical data in this case can be directly compared with exponentials 
in $r$, without any power prefactor. The connected wall-wall correlator in the 
$x$-direction is defined as
\be
G(x)=\frac{1}{L}\sum_{x_0} \langle w(x_0+x)^\dagger w(x_0) \rangle -\langle 
w(x_0+x)^\dagger \rangle \langle w(x_0) \rangle \;,
\label{corr}
\ee
where
\be
w(x)=\frac{1}{L^2}\sum_{y,z} s_{\{x,y,z\}}
\label{wall}
\ee 
represents the spin average over the ``wall'' at the coordinate $x$.

The general behavior for the function $G(x)$ is
\be
G(x)= \sum_n a_n e^{-m_n x} \;,
\label{corr-funct}
\ee
where $m_0$ is the fundamental mass, while $m_1$, $m_2$, ... are higher masses
with the same angular momentum and parity (0$^+$) quantum numbers 
of the fundamental mass.
Mass excitations in channels different from 0$^+$ can be determined by a 
suitable redefinition of the wall average~(\ref{wall}). The fundamental mass
in a definite channel can be extracted from wall-wall correlators by looking for 
a plateau of the effective mass,
\be
m_{\mbox{\footnotesize eff}}(x)= \ln \frac{G(x)}{G(x-1)} \;,
\ee
at large distances. Excited masses in the given channel can be found by the variational 
method~\cite{Kronfeld,Luscher-Wolff}, which consists in defining several
wall-averaged operators with the given quantum numbers and building the
matrix of cross-correlations between them. The eigenvalues of this matrix 
are single exponentials of masses in the given channel. The possibility to
determine masses above the fundamental one in a given channel relies on the
ability to define operators which have a large overlap with the excited states.
Usually, these operators are defined in order to probe different length scales.
In the present work we considered only the 0$^+$ and the 2$^+$ channels; the 
local variables to be wall-averaged as in~(\ref{wall}) have been defined 
in the following way:
\begin{eqnarray}
s^{0^+}_{\{x,y,z\}}(n)&=&s_{\{x,y,z\}}(s_{\{x,y+n,z\}}+s_{\{x,y,z+n\}}) \;, 
\nonumber \\
s^{2^+}_{\{x,y,z\}}(n)&=&s_{\{x,y,z\}}(s_{\{x,y+n,z\}}-s_{\{x,y,z+n\}}) \;.
\label{var}
\end{eqnarray}
Typically, we have used about ten operators in each channel, with the largest 
$n$ almost reaching $L$. 
We can anticipate that we have not been able to extract masses 
higher than the fundamental one in any of the two channels. However, the use of
the variational method has improved considerably the evaluation of the fundamental
2$^+$ mass. Needless to say that the determination of the fundamental mass
in the 0$^+$ channel by the definition~(\ref{var}) turned out to be in perfect
agreement with that from~(\ref{corr}).

A few years ago it has been proposed~\cite{Caselle:1999tm} that ratios between 
massive excitations in the broken phase are universal quantities, i.e. mass
ratios must be equal in theories where the same mechanism of symmetry breaking 
at the transition is active. One of the motivations of the present
 work is to check whether this  
statement holds also for the broken phase of the Ising theory~(\ref{ising}) near the
critical endpoint $P_c$ of the 3$d$ 3-state Potts model. This would 
provide us with a non-trivial check of validity of the conjecture of
universal mass spectrum. 

For the 3$d$ Ising universality class, a definite prediction exists for the ratio 
between the fundamental 0$^+$ and 2$^+$ masses, 
$m_{2^+}/m_{0^+}$~\cite{Caselle:1999tm}. We can estimate the same ratio near 
the Ising endpoint of the 3$d$ 3-state Potts model and verify if it is consistent 
with this prediction. The procedure to determine the fundamental masses in the
two channels of interest is the same outlined above, with the only 
difference that we need to use the correct local order parameter to build 
correlators. We have defined this local variable, $\tilde{m}_i$, in such a way that
it reconstructs the global magnetization operator 
$\tilde{M}$ after summation over the whole lattice:
\be
\tilde{m_i} = \delta_{\sigma_i \sigma_h} + \frac{s}{2} \sum_{\hat{\mu}} 
\delta_{\sigma_i \sigma_{i+\hat{\mu}}} \;.
\ee
Indeed, it is easy to see that $\tilde{M}_i = \sum_i \tilde{m}_i$.

The second aim of the present work is to study mass ratios 
also in the broken phase of the 3$d$ 3-state Potts model at zero magnetic field near 
the transition point. 
It is known that this transition is weakly first order, a signature of this
being the fact that the correlation length $\xi$, though finite at the
transition point, becomes much larger than the lattice spacing.
This fact could open the door to universality effects, such as the above-mentioned
conjecture on mass ratios. If this claim is true, an important consequence would be
that mass ratios in the broken phase near criticality of the 3$d$ 3-state Potts model
at zero magnetic field should reproduce the corresponding ratios in the
phase of broken Z(3) symmetry of the (3+1)$d$ SU(3) pure gauge theory at finite
temperature. This is the deconfined phase of the SU(3) gauge theory and the
counterparts of the massive excitations of the spin model are the so-called
{\it screening} masses, i.e. the inverse decay lengths of the Yukawa-like
interaction potential between static color sources.

As a probe for the check of consistency of this scenario, we can consider
again the mass ratio $m_{2^+}/m_{0^+}$. If it will turn out that its value
near the transition coupling $\beta_t$ in the broken phase of the 3$d$ 3-state
Potts model is compatible with the corresponding ratio of screening masses in the
deconfined phase near transition of (3+1)$d$ SU(3) at finite temperature, this
could be taken as an indication that the conjecture of universal spectrum
can be extended also to the case of weakly first order transition.

Summarizing, we have to calculate the $m_{2^+}/m_{0^+}$ ratio in two regions of 
interest in the temperature - magnetic field phase diagram: (a) in the broken phase 
near the critical Ising endpoint, (b) in the broken phase near the transition point 
of the model in absence of magnetic field. In the case (a), we have to compare the
result with the predicted value from universality in the 3$d$ Ising class, 
given in Ref.~\cite{Caselle:1999tm}; in the case (b), our determination is to be
compared with the corresponding one in the SU(3) pure gauge 
theory~\cite{Falcone-Fiore-Gravina-Papa}. 

\section{Numerical results}
\label{results}

We have performed numerical Monte Carlo simulations of the 3$d$ 3-state Potts 
model using a cluster algorithm~\cite{Swendsen-Wang,Kasteleyn-Fortuin} to reduce 
the autocorrelation effects. In order to minimize the finite volume effects, we have 
imposed periodic boundary conditions. Data analysis has been done by the jackknife
method applied to bins of different lengths. 

For the simulations near the critical endpoint $P_c$ (region (a)) we used lattices
with size $L=70$; near the transition coupling $\beta_t$ at zero magnetic field 
(region (b)) we chose instead $L=48$. In both cases we have seen tunneling between 
degenerate minima near the transition point. This finite volume effect can 
spoil mass measurements in the scaling region and must be treated carefully. 
Depending on the order of the transition, tunneling effects show up
differently and must be removed accordingly.

\subsection{Region (a): critical endpoint}

We have performed simulations on $70^3$ lattices for which the mixing parameters
appearing in~(\ref{mixing}) turn to be $s(L=70)=-0.689(8)$ and 
$r(L=70)=0.690(3)$~\cite{Karsch-Stickan}.

\begin{figure}[tb]
\centering
\includegraphics[width=7.5cm,bb=40 40 700 620]{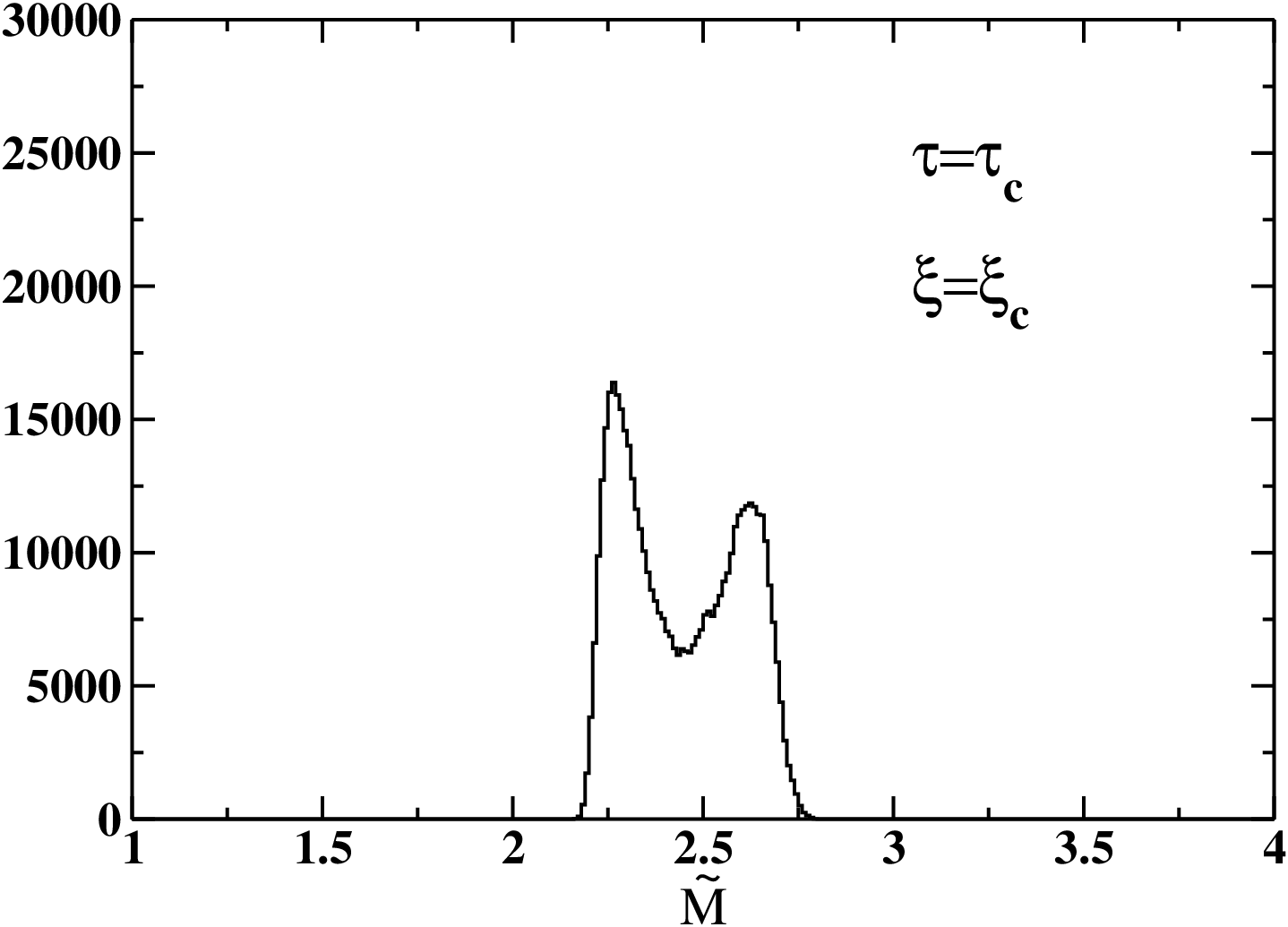}

\includegraphics[width=7.5cm,bb=40 40 700 620]{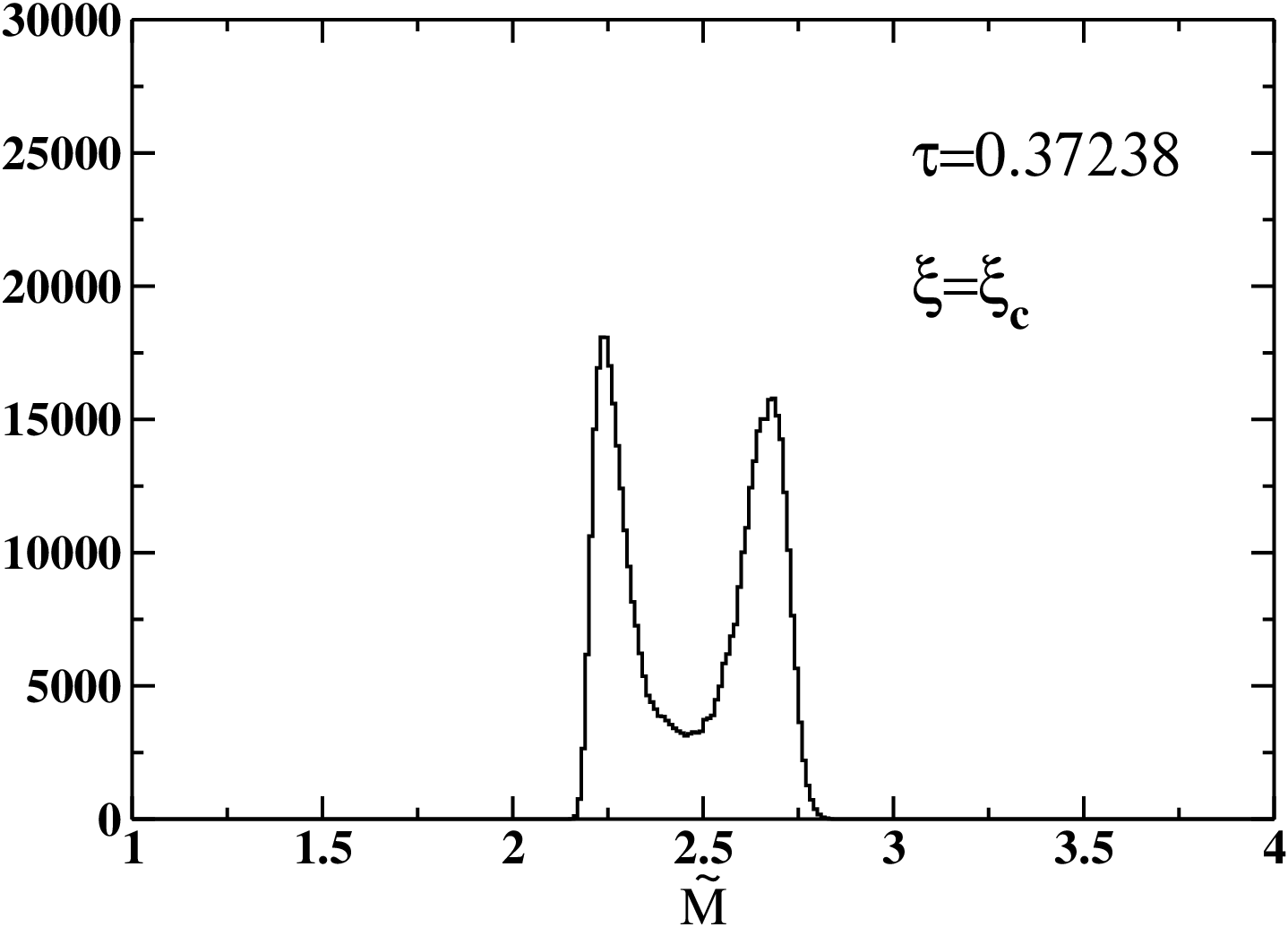}

\includegraphics[width=7.5cm,bb=40 40 700 620]{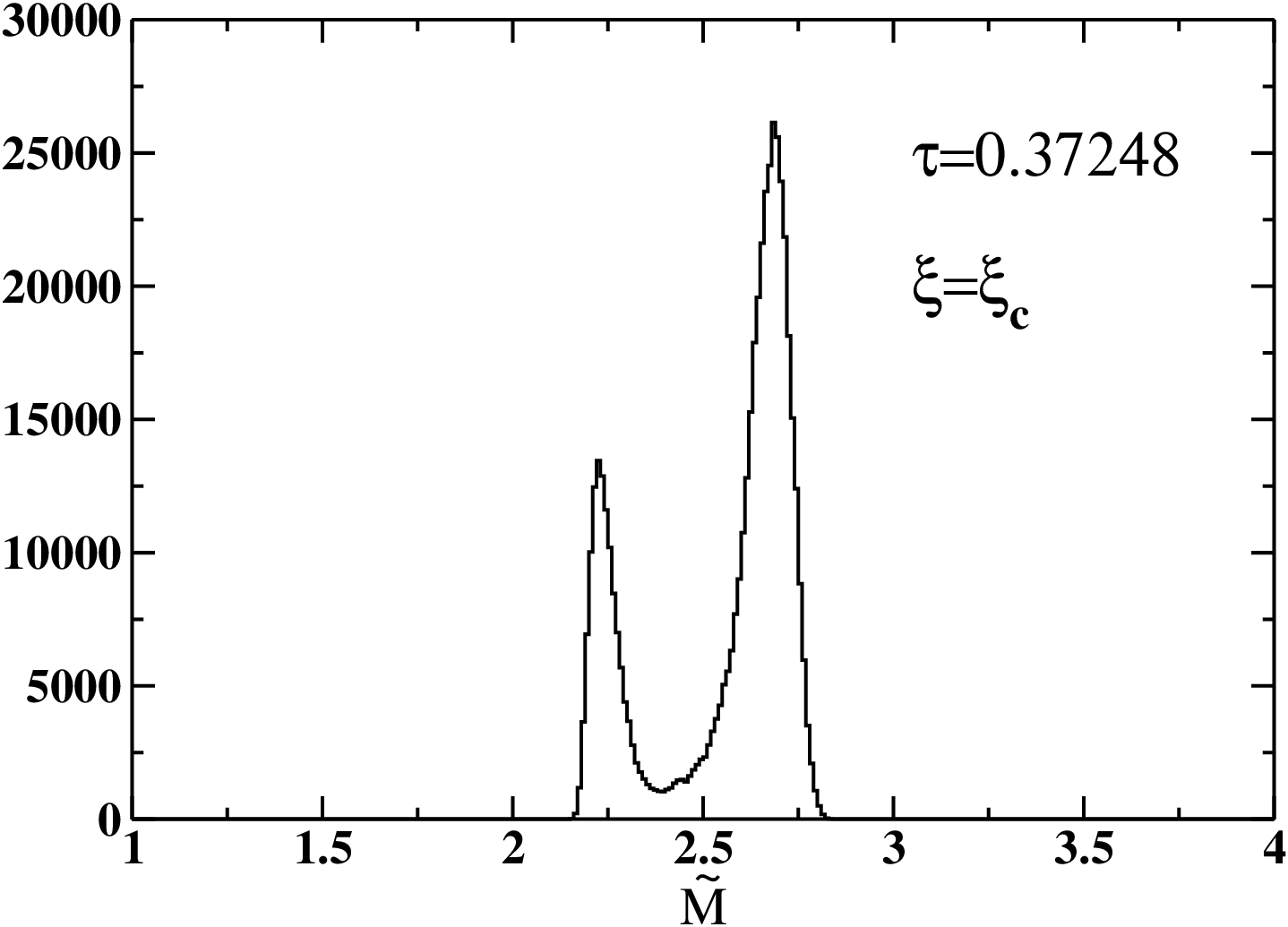}
\caption[]{Distributions of the order parameter $\tilde{M}$ near $P_c$. The
simulations have been done at the points $(\xi_c,\tau)$ 
with $\tau=0.37182\equiv\tau_c$, $\tau$=0.37238 and $\tau$=0.37248. These points 
lie on the first order line starting from $P_c$. The statistics is 500k in all 
cases.}
\label{karsch_peaks} 
\end{figure}

First of all we have considered the distribution of the order parameter 
$\tilde{M}$ in the broken phase of the Ising theory~(\ref{ising}) near $P_c$.
Fig.~\ref{karsch_peaks} shows the structure, typical for a second-order
phase transition, with two peaks corresponding to the two degenerate 
minima in the broken phase which separate while moving away from $P_c$ along the 
first order Ising critical line. This double-peak structure is the signal of tunneling.
The last plot in Fig.~\ref{karsch_peaks}, obtained for Ising couplings $\xi=\xi_c$ 
and $\tau$=0.37248, shows two almost completely separated peaks. 
We have decided to choose this as our working point
and performed here simulations with statistics 200k.

\begin{figure}[tb]
\centering
\includegraphics[width=14cm]{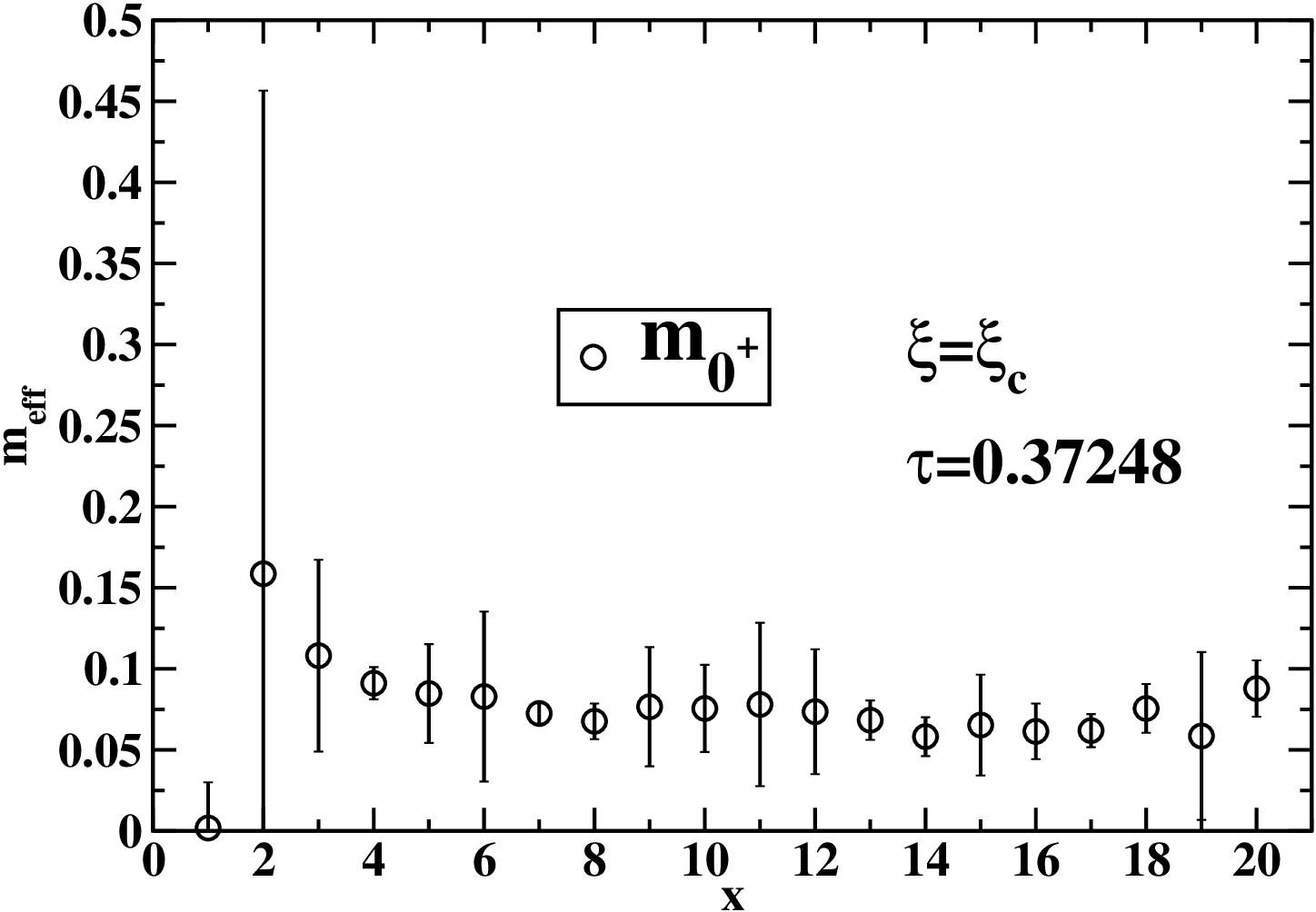}
\caption[]{Effective mass in the 0$^+$ channel as a function of the separation between 
walls on the $(y,z)$ plane at $\xi=\xi_c$ and $\tau$=0.37248, determined from the 
configurations belonging to the ``right-peak'' in the thermal equilibrium ensemble.}
\label{karsch_mass0+_left} 
\end{figure}

\begin{figure}[tb]
\centering
\includegraphics[width=14cm]{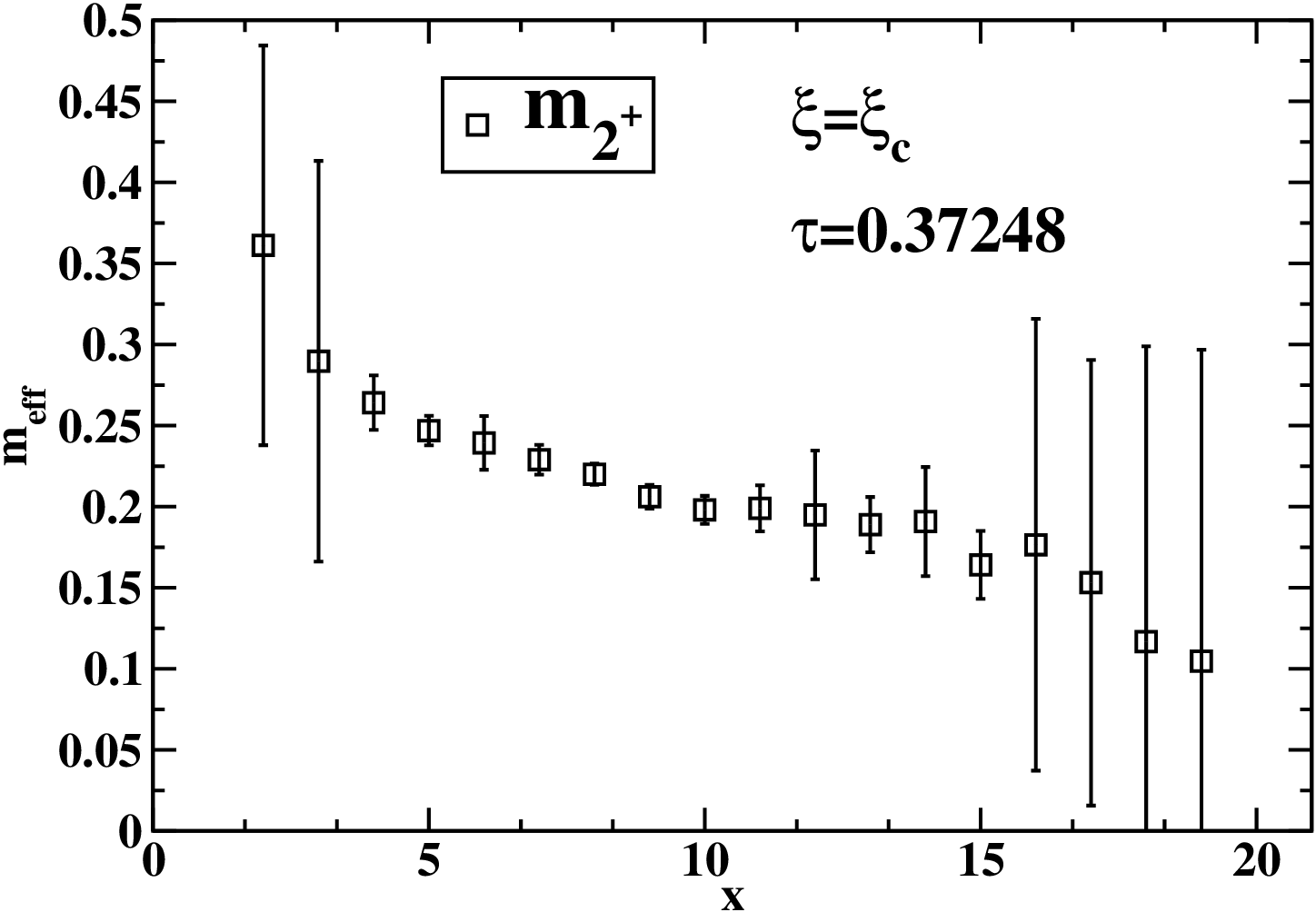}
\caption[]{Effective mass in the 2$^+$ channel as a function of the separation between 
walls on the $(y,z)$ plane at $\xi=\xi_c$ and $\tau$=0.37248, determined from the 
configurations belonging to the ``right-peak'' in the thermal equilibrium ensemble.}
\label{karsch_mass2+_left} 
\end{figure}

We have removed here tunneling effects by the brute force method of analyzing 
separately data belonging to each peak. 
In Figs.~\ref{karsch_mass0+_left} and ~\ref{karsch_mass2+_left} 
we show the behavior of the effective masses in the 0$^+$ and in the 
2$^+$ channels at $\xi=\xi_c$ and $\tau$=0.37248, as functions of the separation 
between walls, for the configurations in the ``right-peak''. Similar plots have been obtained 
for the configurations in the ``left-peak''. In each case, the {\it plateau mass}
is taken as the effective mass (with its error) belonging to the {\it plateau} 
and having the minimal uncertainty. We define {\it plateau} the largest set of consecutive
data points, consistent with each other within 1$\sigma$.
This procedure is more conservative than identifying the plateau mass and its error
as the results of a fit with a constant on the effective masses $m_{\mbox{\footnotesize eff}}(x)$,
for large enough $x$. We have found

\vspace{0.5cm}

``right-peak'' (statistics 115k)
\begin{equation}
am_{0^+}= 0.0725(63) \;, \;\;\;\;\;
am_{2^+}= 0.1981(87) \;, \;\;\;\;\;
\frac{m_{2^+}}{m_{0^+}} = 2.73(36) \;;
\end{equation}

``left-peak'' (statistics 85k)
\begin{equation}
am_{0^+}= 0.0714(40) \;, \;\;\;\;\;
am_{2^+}= 0.1959(80) \;, \;\;\;\;\;
\frac{m_{2^+}}{m_{0^+}} = 2.74(27) \;.
\end{equation}
The uncertainty on the mass ratios has been determined by usual propagation of errors.
The two determinations of the masses and, therefore, of the mass ratios are 
consistent, as expected. Moreover, they are
compatible with the value of the 3$d$ Ising class~\cite{Caselle:1999tm}, 
$m_{2^+}/m_{0^+}= 2.59(4)$.

\subsection{Region (b): transition point at zero magnetic field}

We have performed simulations on $48^3$ lattices for several values of the
coupling $\beta$ in the broken phase of the 3$d$ 3-state Potts model at zero magnetic
field. A summary of the Monte Carlo simulations is presented in Table~\ref{masses}.

\begin{figure}[tb]
\centering
\includegraphics[width=9cm]{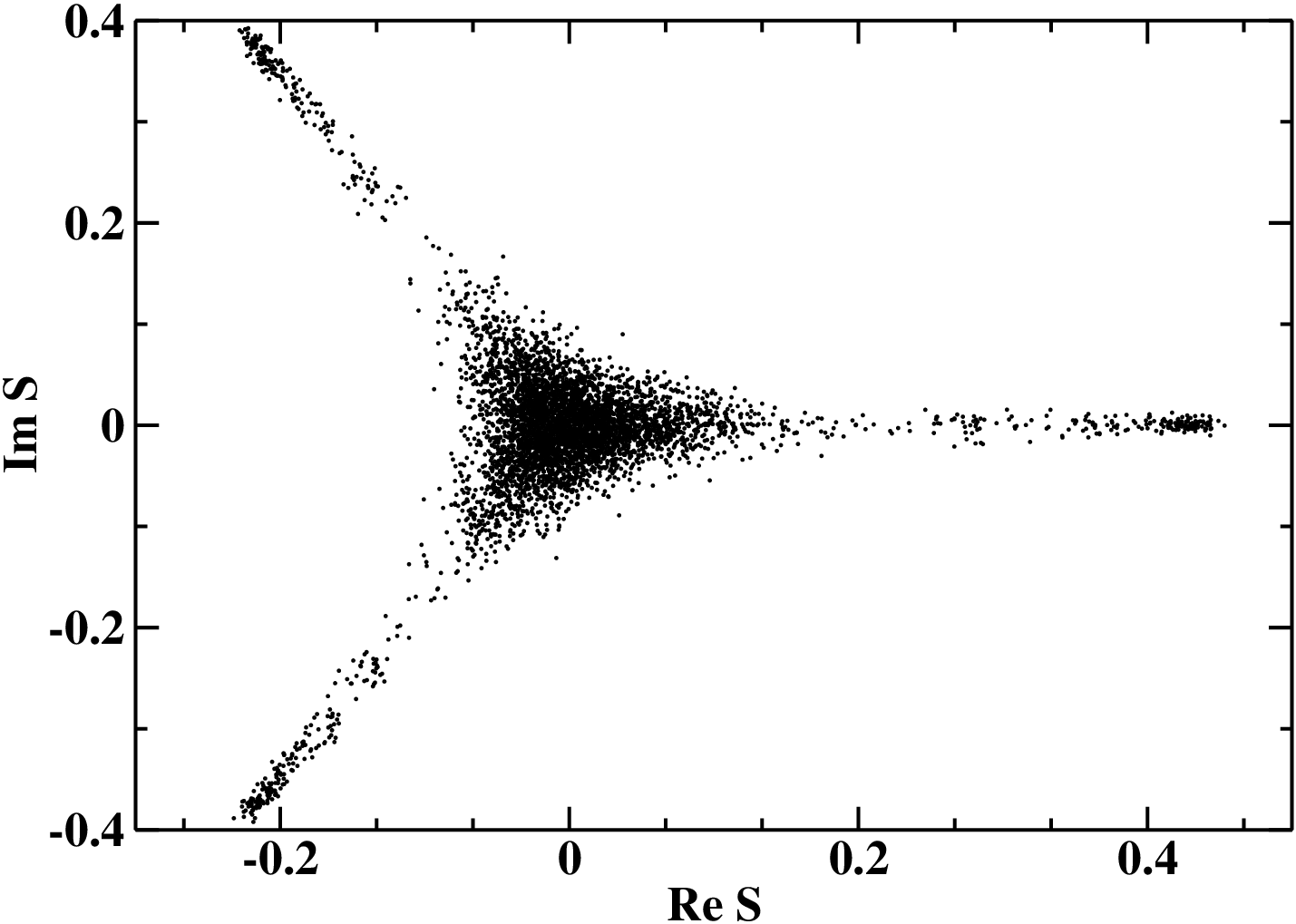}
\caption[]{Scatter plot of the complex variable $S$ at the transition
point $\beta_t$=0.550538~\cite{Gavai-Karsch-Petersson} at zero magnetic field. 
Both the symmetric and the broken phases are present.}
\label{scatter_trans}
\end{figure}

Close enough to $\beta_t(L=48)$=0.550538, determined in 
Ref.~\cite{Gavai-Karsch-Petersson}, the scatter plot of the complex 
parameter $S$ shows the coexistence of the symmetric phase 
(points around (0,0) in the Im$\,S$ - Re$\,S$ plane
in Fig.~\ref{scatter_trans}) and of the broken phase (points around the three 
roots of the identity in Fig.~\ref{scatter_trans}). Notice that the peaks
in the distribution of points in Fig.~\ref{scatter_trans} are well separated,
as it must be for first order dynamics. In the thermodynamic limit this would
occur only at the transition point; at finite volume, tunneling between broken minima 
and the symmetric phase occurs in a small region around the transition. The amplitude 
of this region decreases with the volume. For $L$=48 it is of the order of $10^{-4}$ 
in $\beta$~\cite{Gavai-Karsch-Petersson}.

\begin{figure}[tb]
\centering
\includegraphics[width=9cm]{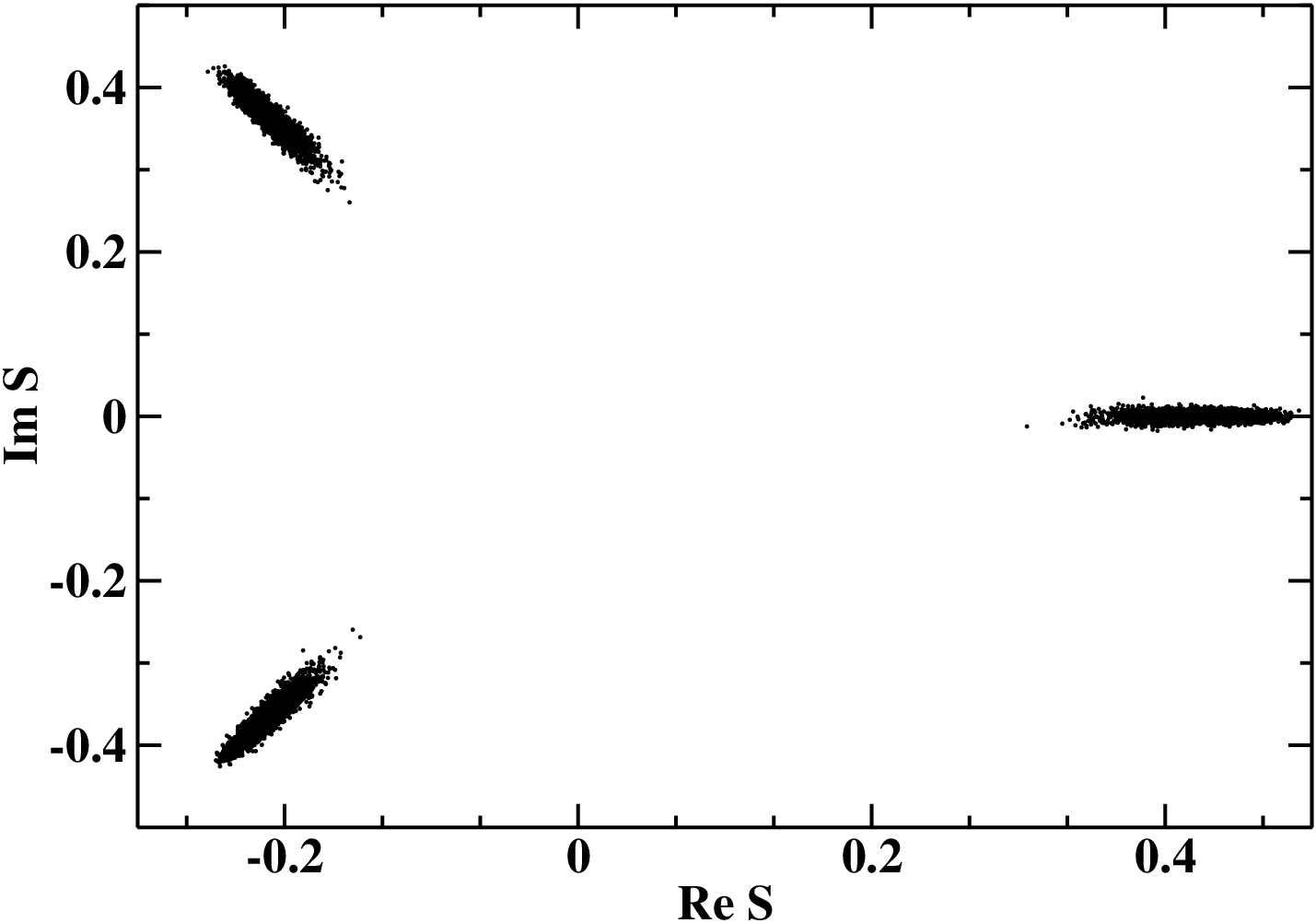}

\vspace{2cm}

\includegraphics[width=9cm]{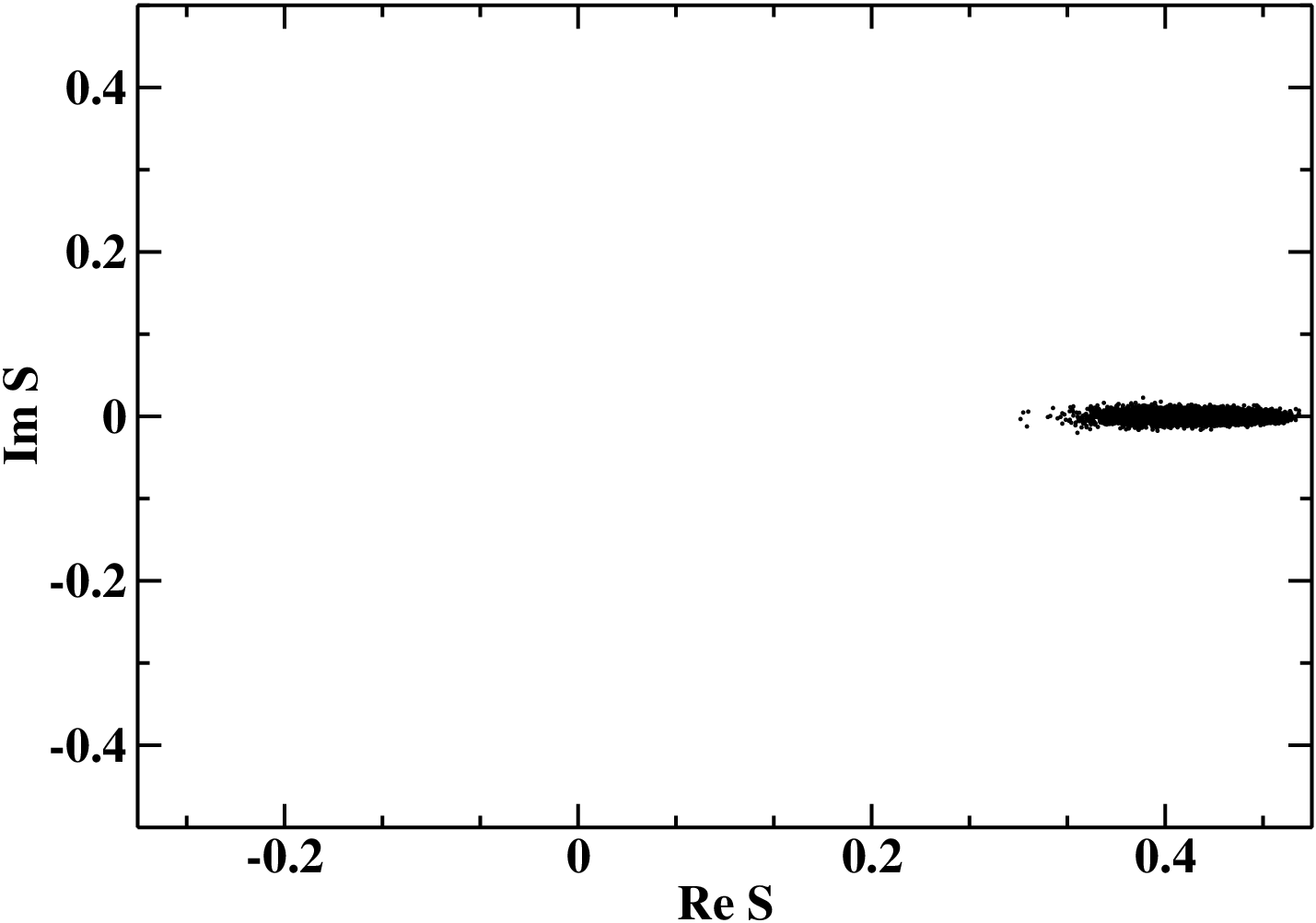}
\caption[]{(Top) Typical scatter plot of the complex variable $S$ for
$\beta$ larger than 0.5508 and zero magnetic field on a lattice 48$^3$. 
There are no states in the symmetric phase,
but tunneling survives between the three broken minima.\newline
(Bottom) Same as (Top) with the tunneling between broken minima removed by 
the ``rotation'' to the real phase.}
\label{0.5508_rotation}
\end{figure}

\begin{figure}[tb]
\centering
\includegraphics[width=14cm]{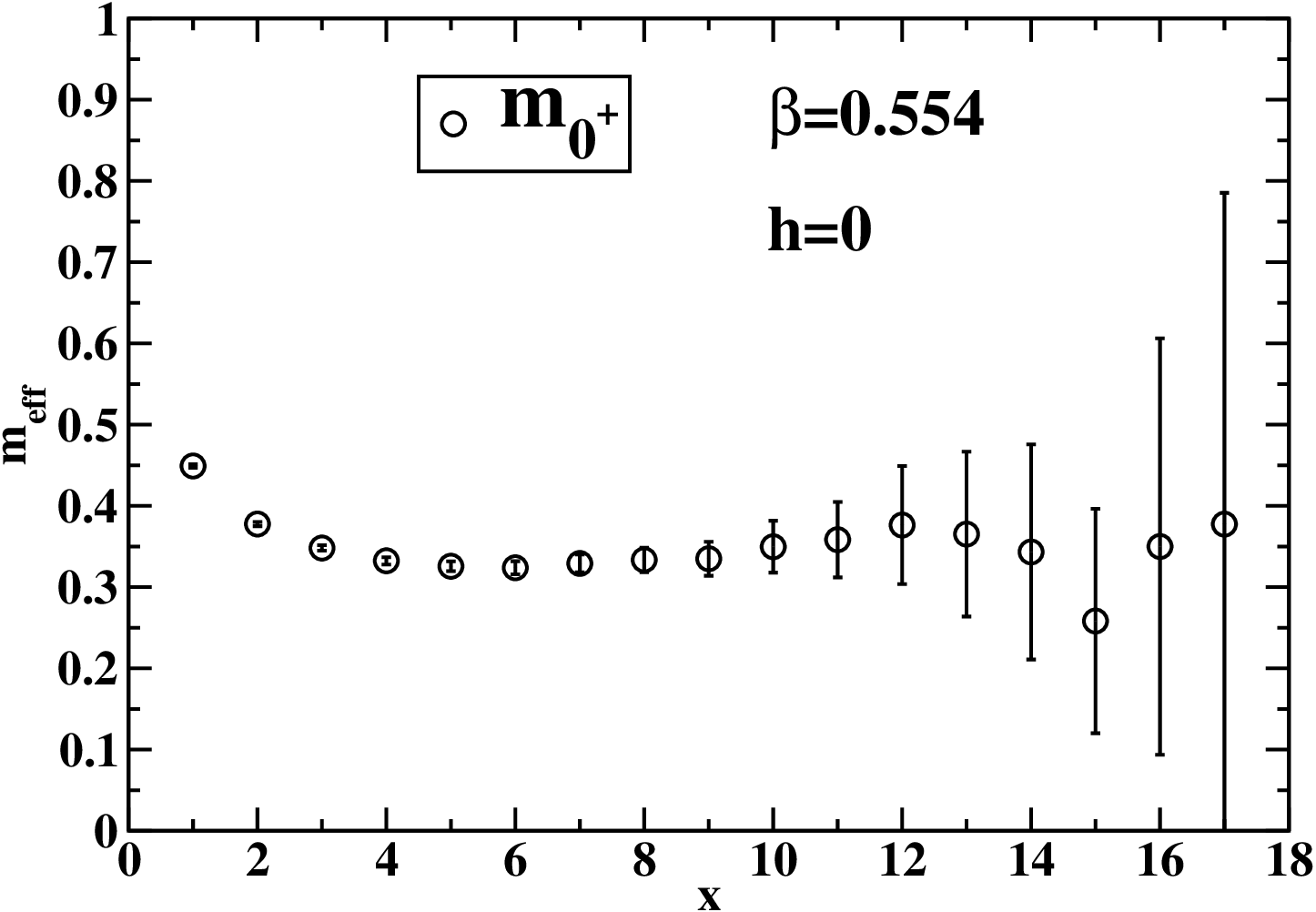}
\caption[]{Effective mass in the 0$^+$ channel as a function of the separation between 
walls on the $(y,z)$ plane at $\beta=0.554$ and $h=0$.}
\label{mass0+_beta=0.554} 
\end{figure}

\begin{figure}[tb]
\centering
\includegraphics[width=14cm]{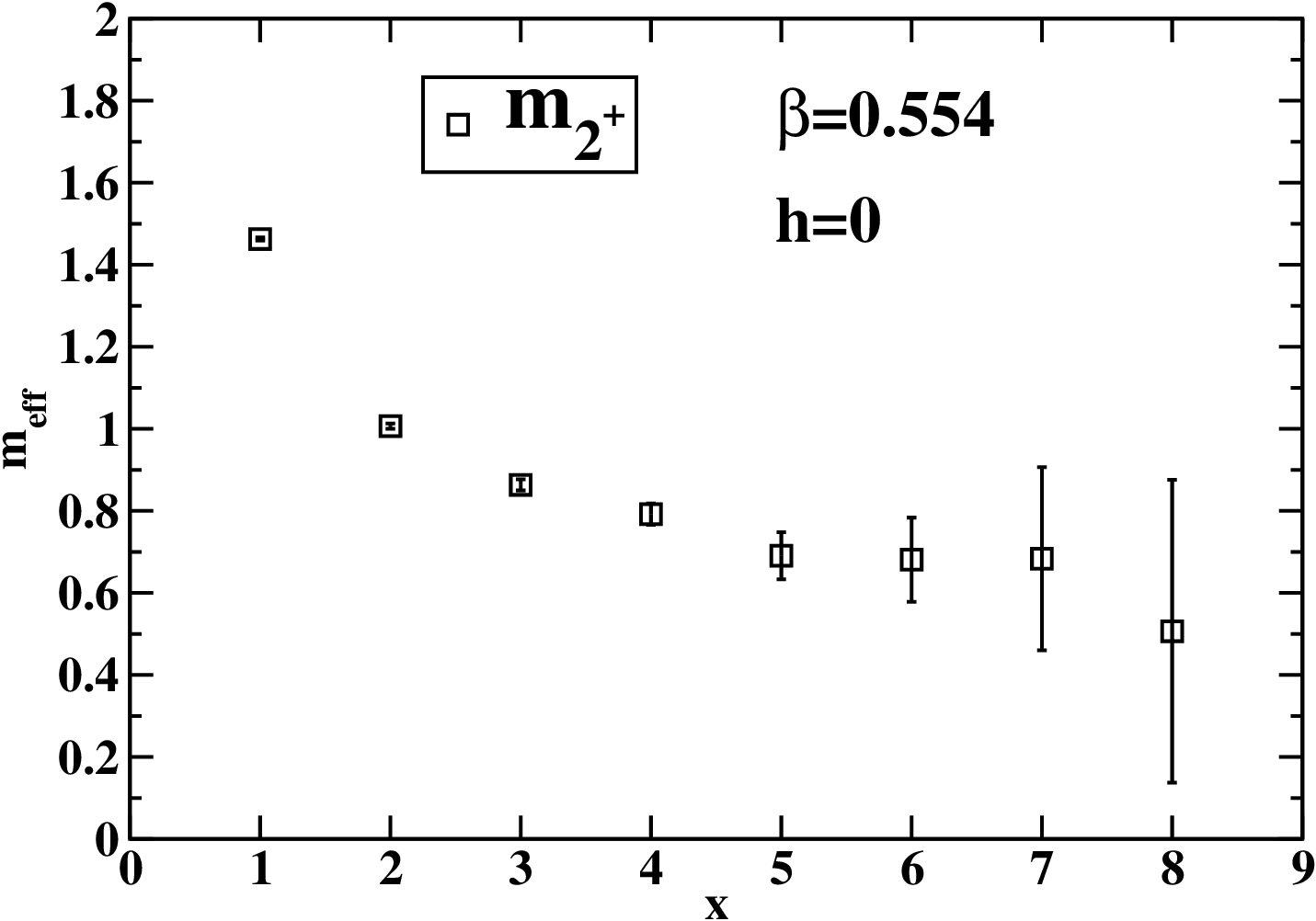}
\caption[]{Effective mass in the 2$^+$ channel as a function of the separation between 
walls on the $(y,z)$ plane at $\beta=0.554$ and $h=0$.}
\label{mass2+_beta=0.554} 
\end{figure}

\begin{figure}[tb]
\centering 
\includegraphics[width=14cm]{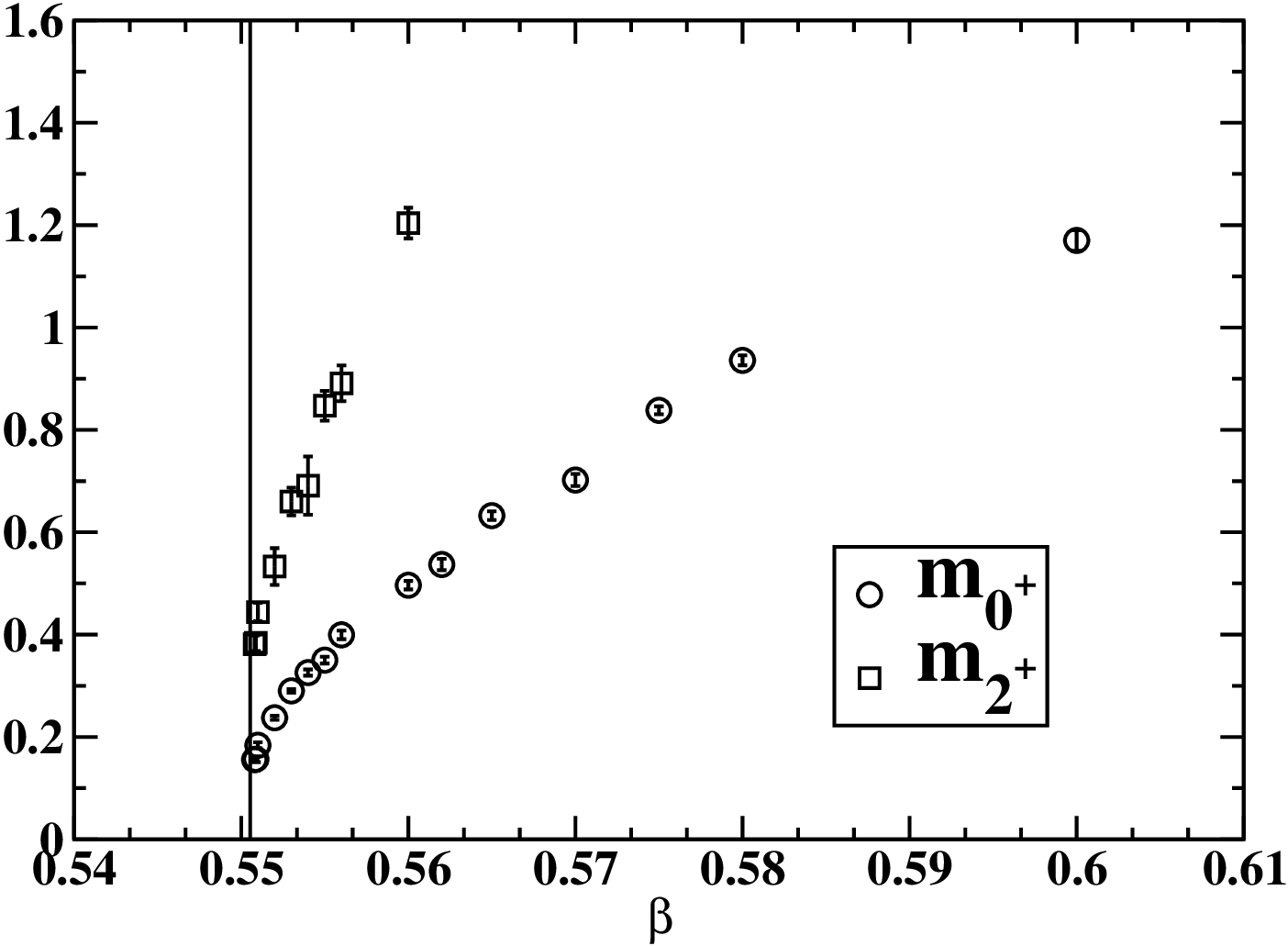}
\caption[]{Fundamental masses in the $0^+$ and $2^+$ channels as functions of $\beta$, 
in the broken phase near $T_t$ (vertical line).}
\label{masses_0_2}
\end{figure}

Moving away from $\beta_t$ the symmetric phase becomes less and less important, 
up to disappearing. For $\beta$=0.5508 there are only the three broken minima,
as shown in Fig.~\ref{0.5508_rotation}(top). For larger $\beta$ values the tunneling 
between broken minima survives. However, the three peaks are totally separated and 
it is therefore possible to ``rotate'' unambiguously all of them to the real sector 
(see Fig.~\ref{0.5508_rotation}(bottom)). Working only in one sector allows us to 
optimize statistics. With this approach we have calculated the fundamental masses 
in the $0^+$ and $2^+$ channels for several $\beta$ values, up to 0.60 
($(T_t-T)/T_t\sim 0.08$). In Figs.~\ref{mass0+_beta=0.554} and ~\ref{mass2+_beta=0.554}
we show the behavior of the effective masses in the 0$^+$ and in the 
2$^+$ channel at $\beta=0.554$, as functions of the separation between walls. Similar
plots have been obtained for the other $\beta$ values. The plateau mass values have been 
determined as described in the previous Subsection. In Table~\ref{masses} we present our 
results, whereas in Fig.~\ref{masses_0_2} we have plotted the behavior of $m_{0^+}$
and $m_{2^+}$ versus $\beta$.

\begin{table}[htb]
\centering
\caption[]{Fundamental masses in lattice units in the $0^+$ and $2^+$ channels 
and their ratio for $\beta$ values in the broken phase near the weakly first 
order transition at $L$=48. The statistics of each simulation is also given.}

\vskip 0.4cm

\begin{tabular}{|l|l|c|c|c|}
\hline
 $\beta$ & $m_{0^+}$ & $m_{2^+}$ & $m_{2^+}$/$m_{0^+}$ & statistics \\
\hline
0.5508   & 0.1556(36) & 0.381(17) & 2.45(17) & 300k \\
0.550875 & 0.1565(56) & 0.384(16) & 2.45(19) & 200k \\
0.551    & 0.1837(59) & 0.444(18) & 2.42(18) & 100k \\
0.552    & 0.2375(42) & 0.533(36) & 2.24(19) & 100k \\
0.553    & 0.2900(34) & 0.660(27) & 2.28(12) & 200k \\
0.554    & 0.3258(60) & 0.691(57) & 2.12(21) & 100k \\
0.555    & 0.3502(67) & 0.847(29) & 2.42(13) & 200k \\
0.556    & 0.3996(85) & 0.891(35) & 2.23(14) & 200k \\
0.56     & 0.4965(83) & 1.204(30) & 2.43(10) & 200k \\
0.562    & 0.537(11)  & -         & -        & 200k \\ 
0.565    & 0.6324(85) & -         & -        & 200k \\
0.57     & 0.702(12)  & -         & -        & 200k \\ 
0.575    & 0.8381(78) & -         & -        & 200k \\
0.58     & 0.9358(97) & -         & -        & 200k \\
0.60     & 1.170(20)  & -         & -        & 200k \\
\hline
\end{tabular}
\label{masses}
\end{table} 

\begin{figure}[tb]
\centering
\includegraphics[width=14cm]{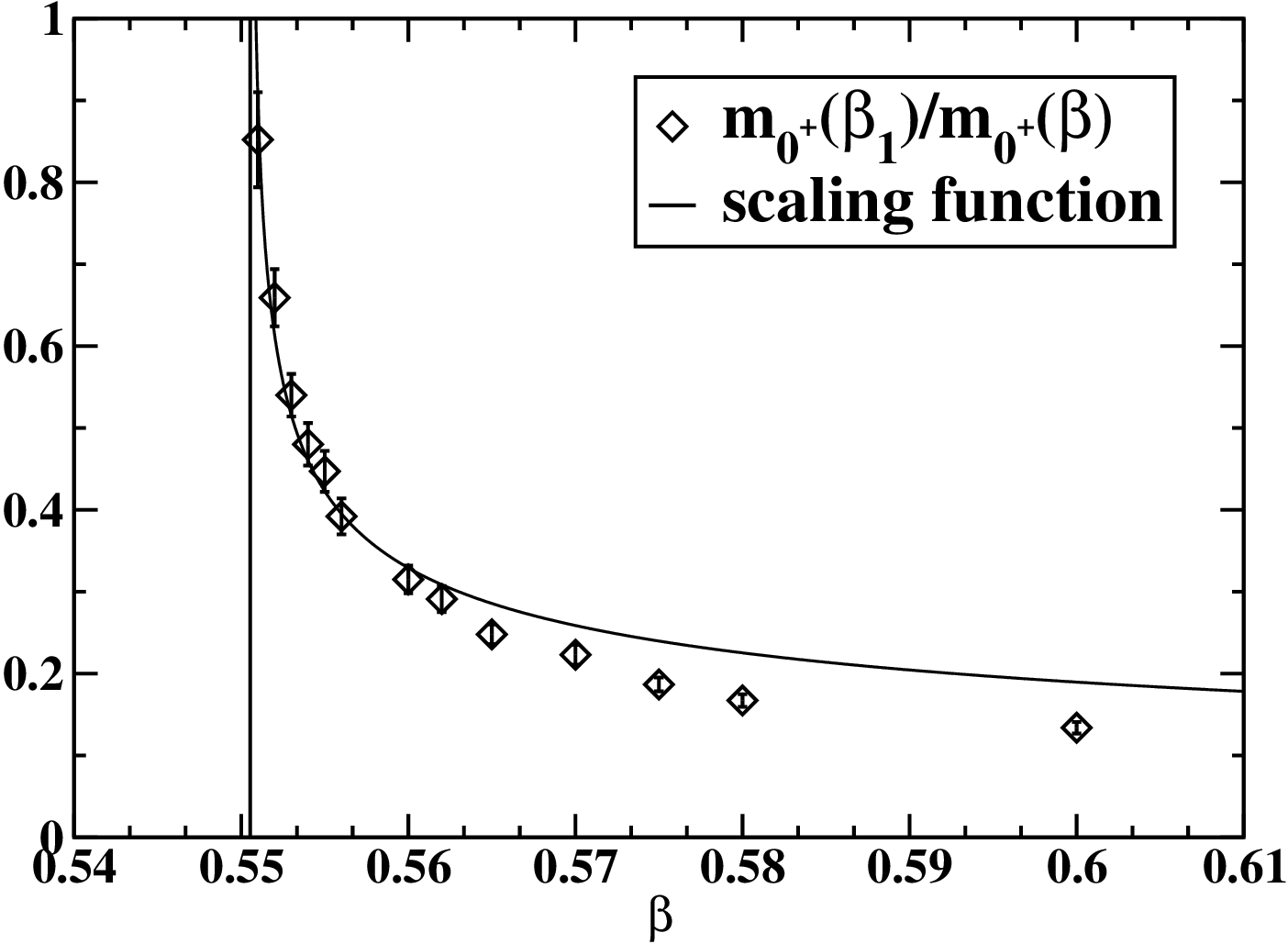}
\caption[]{Comparison between the scaling function $[(\beta_1-\beta_t)/(\beta-\beta_t)]^{1/3}$
and the mass ratio $m_{0^+}(\beta_1)/m_{0^+}(\beta)$ for varying $\beta$, with
$\beta_1$=0.550875.} 
\label{scaling}
\end{figure}

Since for first order transitions the critical exponent of the correlation length is 
$\nu=1/3$~\cite{Fisher-Berker}, the scaling law
\be
\label{scal_rel}
\Bigg( \frac{\beta_1-\beta_t}{\beta_2-\beta_t} \Bigg)^{\nu} \sim 
\frac{m_{0^+}(\beta_1)}{m_{0^+}(\beta_2)} 
\ee
must hold in the scaling region. Here $m_{0^+}(\beta_1)$ and 
$m_{0^+}(\beta_2)$ are the fundamental masses in the $0^+$ channel at $\beta_1$ 
and $\beta_2$, respectively. Comparing the scaling function with the ratios 
determined from simulations, we can estimate how large is the scaling region. 
Fig.~\ref{scaling} shows that the scaling region extends at least up to 
$\beta=0.562$ ($(T_t-T)/T_t\sim 0.02$).

\begin{figure}[tb]
\centering
\includegraphics[width=14cm]{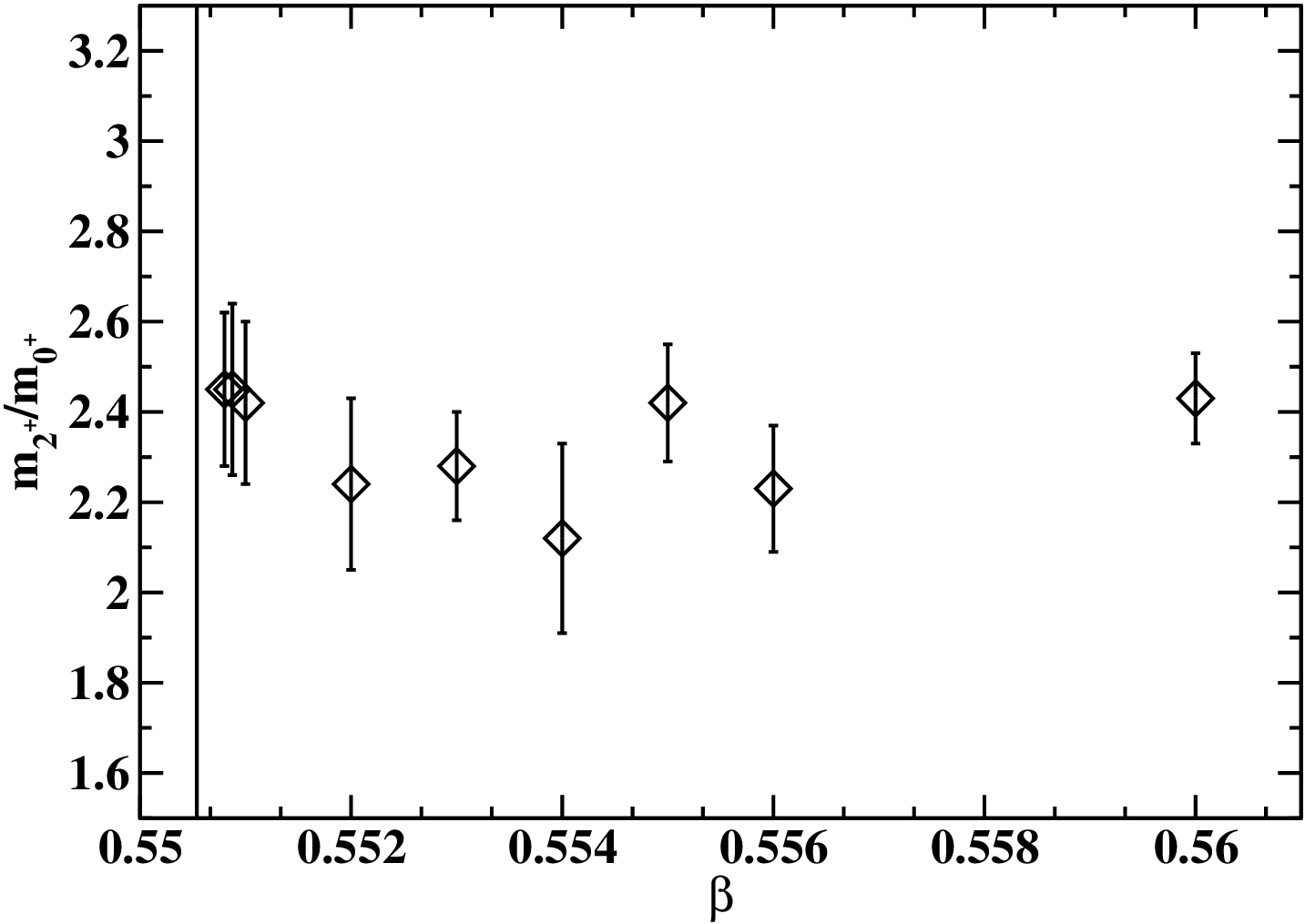}
\caption[]{$m_{2^+}(\beta)/m_{0^+}(\beta)$ for $\beta$ varying in the scaling region.}
\label{ratio}
\end{figure}

We have determined the ratio $m_{2^+}/m_{0^+}$ for several $\beta$ values
in the region $[\beta_t,0.56]$; the results are presented in Table~\ref{masses} 
and plotted in Fig.~\ref{ratio}. This ratio remains practically constant in the 
considered region, thus suggesting that the correlation length $\xi_2$ associated 
to the channel $2^+$ ($\xi_2=1/m_{2^+}$) scales in the same way of the fundamental one 
($\xi_0=1/m_{0^+}$). We can take as our estimation of the mass ratio the value
\be
\frac{m_{2^+}}{m_{0^+}} = 2.43(10) \;,
\label{ratioval}
\ee
determined, as discussed above, by taking value and error of the point
with the smallest error belonging to the plateau ($\beta=0.56$, see
Fig.~\ref{ratio}). A fit of the data with a constant gives for this ratio
the value 2.353(49) with a $\chi^2$/d.o.f.= 0.54.

The result given in~(\ref{ratioval}) is to be taken as a prediction for 
the corresponding ratio of the screening masses in the broken phase of the
(3+1)$d$ SU(3) pure gauge theory at finite temperature.

\section{Conclusions and outlook}

In this work we have studied massive excitations of the 3$d$ 3-state Potts model
near the Ising critical point on the temperature - magnetic field phase diagram
and near the transition point at zero magnetic field.

We have found evidence that the mass ratio $m_{2+}/m_{0+}$ near the 
Ising critical point is compatible with the prediction from universality, thus
supporting the conjecture of universal spectrum.

In the broken phase of the scaling region near the transition in absence 
of the external source, we have found $m_{2+}/m_{0+}$=2.43(10). This result 
is to be compared with the corresponding ratio between screening masses of (3+1)$d$
SU(3) pure gauge theory at finite temperature in the broken phase 
near the deconfinement temperature~\cite{Falcone-Fiore-Gravina-Papa}. An
agreement between these determinations would represent an indication that
universality of the mass spectrum holds also for weakly first order transitions.

The present analysis could be extended with the numerical determination of 
higher masses in the $0^+$ and $2^+$ channels and with masses in other 
channels, in order to carry out a more systematic study of universality effects.
Such an extension would be certainly well-motivated in case of a positive answer
of the comparison with the SU(3) gauge theory.

\end{document}